\documentstyle[12pt]{article}
\def\CC{{\rm \kern.24em \vrule width.02em height1.4ex
depth-.05ex \kern-.26em C}}
\def \qed {\hfill \vrule height6pt width 6pt depth 0pt}

\begin{document}

\title{A completely entangled subspace of maximal dimension}

\author{\large{B. V. Rajarama Bhat}\\
 Indian Statistical Institute,\\
R. V. College Post, Bangalore 560059, India.\\
 E-mail: {\sl bhat@isibang.ac.in} }

\date{September 1, 2004}

\maketitle

\begin{abstract}
Consider a tensor product
${\cal H}= {\cal H}_1\otimes {\cal H}_2\otimes\cdots \otimes {\cal H}_k$ of finite dimensional
Hilbert spaces with  dimension of ${\cal H}_i=d_i, 1\leq i\leq k$. Then the maximum dimension
possible for a subspace of ${\cal H}$ with no non-zero product vector is
known to be
$d_1d_2\ldots  d_k-(d_1+d_2+\cdots +d_k)+k-1.$ We obtain an explicit example
of a subspace of this kind. We determine the set of product vectors in its
orthogonal complement and show that it has the minimum dimension possible for
an unextendible product basis of not necessarily orthogonal product vectors.

\end{abstract}



Let ${\cal H}_i, ~ l \leq i \leq k$ be finite dimensional complex Hilbert spaces
with $\dim({\cal H}_i)$ $=d_i,~~1\leq i\leq k$.  To avoid trivialities we assume that
$d_i \geq 2$ for every $i$.  A state $\rho$ on the tensor product space ${\cal H}
={\cal H}_1 \otimes \ldots \otimes {\cal H}_k$ is said to be {\it separable} if it
is a convex combination of product states. Otherwise it is said to be {\it
entangled}. Entangled states play an important role in quantum computation and
quantum coding theories.  However in general it is not easy to decide whether a
state is entangled or not.  It is a result of Horodecki and Horodecki \cite{HH} that if a state is
separable then its support is spanned by product vectors (vectors of the form
$y_1\otimes \ldots \otimes y_k)$ in the support. In this context K.R.
Parthasarathy \cite{Pa} defines a subspace ${\cal S} \subset {\cal H}$ as {\em completely
entangled\/} if it does not contain a single non-zero product vector.  This in
particular means that any state $\rho$ with its support in ${\cal S}$ is automatically
entangled.  So the construction of entangled states becomes easy if we have explicit
completely entangled subspaces at our disposal.

The notion of completely entangled subspaces is very closely connected with notions of
unextendible product bases and uncompletable product bases introduced earlier by
D. P. Vincenzo et al. \cite{VM}. We say that a linearly independent subset $B$ of prodcut
vectors ${\cal H}$ forms an {\em unextendible product basis} (UPB) if $B^{\bot }$ is completely
entangled. If in addition the vectors in $B$ are mutually orthogonal we call it an orthogonal
UPB. It is to be noted that in \cite{VM}  only orthogonal bases are considered. This difference is
crucial (In this respect see also Pittenger \cite{Pi} ). Through some interesting combinatorics and
number theory it was shown by Alon and Lov\'{a}sz \cite{AL} that the minimum dimension
possible for an orthogonal UPB is  $N+1$, where $N=\sum _i(d_i-1),$ unless either (i) $k=2$ and
$2\in \{d_1, d_2\}$; or (ii) $N+1$ is odd and at least one $d_i$ is even. In each of
these two special cases, the minimum dimension possible for an orthogonal UPB
is strictly larger than $N+1$. In contrast to this here in Corollary 4  we show that the
minimum dimension possible for an UPB (of not necessarily orthogonal
vectors) is  always $N+1$. We make use of the basic result from Parthasarathy \cite{Pa}
 that the maximal dimension possible for a completely entangled subspaces is $d_1d_2\ldots d_k-(N+1).$

Parthasarathy obtains an explicit example of a  completely entangled
subspace of maximal dimension and exhibits an orthonormal basis for it only in the very
special case of  $k=2$ and
$d_1=d_2 = n$.  Even in this specail case the basis he obtains is quite complicated.
In this short note we give a very simple construction of a completely entangled
subspace of maximal dimension.  Further we show that the orthogonal complement
of this space is spanned by product vectors.

To  simplify notation we fix an infinite dimensional Hilbert space ${\cal K}$ with
ortho-normal basis $\{e_0, e_1, \ldots,\}$ and identify ${\cal H}_i$ with span$\{e_0,e_1 \ldots, e_{d_{i-1}}\}$, so that for each $i,~~\{e_0, e_1, \ldots,
e_{d_{i-1}}\}$ is an otho-normal basis for ${\cal H}_i$. Now take
$${\cal S} =\mbox{span} \left \{ e_{i_1} \otimes \ldots \otimes e_{i_k} - e_{j_1} \otimes \ldots
\otimes e_{j_k} : \sum \limits ^k_{r=1} i_r =\sum \limits ^k_{r=1}j_r,
\begin{array}{l}  0 \leq i_r
\leq d_r-1,\\1 \leq r \leq k \end{array} \right\}$$ We claim that ${\cal S}$ is a completely
entangled subspace of maximal dimension. In order to analyze the structure of ${\cal S}$
we first identify ${\cal S}^\bot$.  Note that ${\cal H}$ decomposes as
${\cal H} = \bigoplus \limits  ^N_{n=0} {\cal H}^{(n)}$, where
$$ {\cal H}^{(n)} = ~\mbox{span~} \{e_{i_1} \otimes \ldots \otimes e_{i_k}: ~~~ \sum \limits
^k _{r=1} i_r = n\}.$$ Further observe that if ${\cal M}$ is a Hilbert space with
orthonormal basis $\{ x_1,$ $ \ldots, x_p\}$ and ${\cal N} = ~\mbox{span~} \{ x_i - x_j
: 1 \leq i, j \leq p\}$, then $N^\bot = \CC u$, where $u = x_1 + \ldots + x_p$.  Now
it is easy to see that each ${\cal H}^{(n)}$ decomposes as ${\cal H}^{(n)} = {\cal
S}^{(n)} \oplus {\cal T}^{(n)},$ where
$${\cal T}^{(n)} = \CC u_n,~~~~u_n = \sum \limits _{i_1 + \ldots +i_k =n} e_{i_1} \otimes \ldots
\otimes e_{i_k};$$
$${\cal S}^{(n)}=~\mbox{span~}\{e_{i_1} \otimes \ldots \otimes e_{i_k} - e_{j_1} \otimes
\ldots e_{j_k}: ~~ \sum \limits ^k _{r=1} i_r = \sum \limits ^k_{r=1}j_r = n\}.$$
${\cal S}^{(0)}={\cal S}^{(N)}=\{ 0\}$. Moreover ${\cal S} = \bigotimes \limits ^N_{n=0} {\cal S}^{(n)}, ~\mbox{and }~ {\cal
S}^\bot = \bigoplus \limits ^N_{n=0} {\cal T}^{(n)}$.

{\bf Theorem 1:} ${\cal S}$ is a completely entangled subspace of ${\cal H}_1
\otimes \ldots \otimes {\cal H}_k$, of maximal dimension.

{\bf Proof :} As each ${\cal T}^{(n)}$ is one dimensional, we see that $\dim({\cal
S}) = d_1d_2 \ldots d_k - \dim({\cal S}^\bot) = d_1d_2 \ldots d_k -(N+1)$. Now suppose $x_1 \otimes \ldots \otimes x_k$ is a non-zero product
vector in ${\cal S}$. Then for any $\lambda \in \CC, ~~ \langle x_1 \otimes \ldots
\otimes x_k, ~~ \sum \limits ^N_{n=0} \lambda^n u_n \rangle = 0.$ However,\\ $\sum
\limits ^N_{n=0} \lambda^n u_n = \sum \limits ^N _{n=0}~~ \sum \limits _{i_1 +
\ldots +i_k =n} \lambda ^{i_1+ \ldots +i_k} e_{i_1} \otimes \ldots \otimes e_{i_k} =
y^\lambda_1 \otimes \ldots \otimes y^\lambda_k$,
where $y^\lambda_r = \sum \limits ^{d_r-1}_{j=0} \lambda^j e_j, ~~ \mbox{for}~ 1\leq
r \leq k$.
So on one hand we get $\prod \limits ^k_{r=1} \langle x_r, ~~ y^\lambda_r \rangle =0$ for all $\lambda \in \CC.$

On the other hand by basic properties of van der Monde matrix $\{ y^\lambda_r\}$ are
linearly independent for any $d_r$, distinct $\lambda $'s. So $ \# \{ \lambda : \langle x_r, y^\lambda_r \rangle = 0\} \leq
d_r-1 < \infty$ for every $r$. Hence $\# \{ \lambda : \prod
\limits ^k_{r=1} \langle x_r, y^\lambda_r \rangle = 0 \}$ must be finite. This is a
contradiction. Therefore  ${\cal S}$ has no non-zero product vector. \qed

Usually for computational purposes one needs an explicit orthonormal basis for the
entangled space under consideration.
Obtaining simple orthonormal bases for our ${\cal S}$ is not difficult as it suffices
to get orthonormal bases for each ${\cal S}^{(n)}$.  As ${\cal S}^{(n)}$  is the
space orthogonal to $u_n$ in ${\cal H}^{(n)}$, there are many ways we can obtain a
basis for it.  All we need to do is obtain a basis for ${\cal H}^{(n)}$ where $u_n$
is one of the vectors.
For instance we get one such basis by considering non-trivial characters of an
abelian group of order $a_n (d_1, \ldots, d_k)$ (the trivial character gets identified
with $u_n$), where  $a_n (d_1, \ldots, d_k)$ is the dimension of ${\cal H}^{(n)}$.

From the definition of ${\cal H}^{(n)}$ it is clear that
$$a_n (d_1, \ldots , d_k) = \# \{ (i_1, \ldots, i_k):~~ i_1 + \ldots + i_k =n , ~~ 0
\leq i_r \leq d_r-1\}.$$
In other words, $a_n(d_1, \ldots , d_k)$ is the coefficient of $x^n$ in the
polynomial
$$ p(x) = \prod \limits ^k_{r=1} (1 + x + x^2 + \ldots + x^{d_r-1}).$$
In particular if $k=2$, and $d_1 \leq d_2$, then
$$ a_n(d_1,d_2) = \left \{ \begin{array}{lllll} n+1 &~\mbox{if}~& 0 \leq n \leq
d_1-1,\\ d_1 &~\mbox{if}~& d_1-1 < n \leq d_2-1,\\
d_1+d_2 - (n+1) &~\mbox{if}~& d_2-1 < n \leq d_1+d_2-2.
\end{array}\right.\hfill (1.1)$$

And in the case of qubits  i.e., if $d_i \equiv 2, ~~~ 1\leq i \leq k$, then
$$a_n(d_1, \ldots, d_k) =
\left(%
\begin{array}{c}
  k \\
  n \\
\end{array}%
\right) , ~~0 \leq n \leq k.~~~~~~~~~~\hfill (1.2) $$

When $k =2 $, by identifying ${\cal H}_1 \otimes {\cal H}_2$ with $d_1 \times d_2$
matrices in the usual way (identify $e_i\otimes e_j$ with matrix unit $E_{ij}$) and noting that in this identification non-zero product
vectors correspond to rank one matrices we arrive at the following Example.

{\bf Example 1:} Consider the vector space $M_{d_1, d_2}$ of $d_1 \times d_2$ complex
matrices.  If a subspace of $M_{d_1, d_2}$ has no rank one element then it has
dimension atmost $(d_1-1)(d_2-1)$.   One such subspace of maximal dimension is given
by:
$$ \left \{ [a_{ij}] : \sum \limits _{i+j=n} a_{ij} =0~~ \forall n \right \} .$$
Now we determine product vectors of ${\cal S}^\bot$.

{\bf Theorem 2:} The set of product vectors in ${\cal S}^\bot$ is $\{ cz^\lambda : c
\in \CC, ~~\lambda \in \CC \cup \{ \infty\}\},$ where
\begin{eqnarray*}
z^\lambda &=& \bigotimes \limits ^k_{r=1}(e_0 + \lambda e_1 + \ldots + \lambda
^{d_r-1} e_r)~~~ \lambda \in \CC; \\
z^\infty &=& \bigotimes \limits ^k_{r=1} ~e_{d_r-1}.\end{eqnarray*}

{\bf Proof:} Consider arbitrary vectors $y_r = \sum \limits ^{d_{r-1}} _{i=0} a^r_i e_i,
~\mbox{in}~ {\cal H}_r ~\mbox{for}~ 1 \leq r \leq k.$  If $y_1 \otimes y_2 \otimes
\ldots \otimes y_k \in {\cal S}^\bot$, we obtain
$$a^1_{i_1} a^2_{i_2} \ldots a^k_{i_k} = a^1 _{j_1} a^2_{j_2} \ldots
a^k_{j_k}, ~~~~~~~ (1.3)$$
whenever $\sum i_r = \sum j_r.$

Case(i): $a^1_0 a^2_0 \ldots a^k_0 \neq 0.$  In this case we may take $a^r_0 =1,$
for $1 \leq r \leq k $.  Take $\lambda = a^1_1$.  Now (1.3) applied to $k$-tuples of
the form (1, 0, 0, \ldots, 0), (0, 1, 0, \ldots 0), i.e, those $k$-tuples with $
\sum i_r = \sum j_r =1,$~ gives~ $a^r_1 = \lambda, ~~\forall r$.
Then by considerting $k$-tuples with $\sum i_r=\sum j_r=2$ we get $a^r_2=\lambda ^2,
\forall r.$
  Continuing this
way we finally obtain $y_1 \otimes \ldots \otimes y_k = z^ \lambda.$

Case (ii): $a^1_0 a^2_0 \ldots a^k_0 =0.$  Now for $1 \leq r \leq k,$ ~let~ $j_r$ be
the smallest $j$ such that $a^r_j \neq 0$.  So $a^1_{j_1}  a^2_{j_2} \ldots
a^k_{j_k} \neq 0.$  Now if for some $r \neq s,~~ j_r > 0$ and $j_s < d_s-1.$ By
(1.3)  we get $a^1_{j_1} a^2_{j_2} \ldots a^r _{j_r-1} \ldots a^s _{j_s+1}
\ldots a^k _{j_k} = a^1_{j_1} a^2_{j_2} \ldots a^k _{j_k} \neq 0.$ But this is not
possible as $a^r_{j_r-1} = 0.$  In other words if $j_r > 0$, then $j_s = d_s-1$ for
all $s \neq r$.  Interchanging the role of $r$ and $s$ we see that $j_r = d_r-1$ for
all $r$.  This shows that $y_1 \otimes \ldots \otimes y_k = cz^\infty$, for some $c
\in \CC$. \qed

It may be noted that $\lim \limits _{| \lambda| \rightarrow \infty} \frac
{z^\lambda}{(\lambda)^N} = z^\infty$.  This justifies the notation used in Theorem
2.

{\bf Theorem 3:} Any $N+1$ non-zero product vectors in ${\cal S}^\bot$ span whole
of ${\cal
S}^\bot$ and hence form an unextendible product basis.

{\bf Proof:} Consider $B \subset \CC \cup \{\infty\}$, with $\#  B = N+1$.  Let $x =
\sum \limits ^N _{n=0} c_n u_n$, be an arbitrary vector in ${\cal S}^\bot$ such that
$ \langle x, z^\lambda \rangle=0 ~~~\forall \lambda \in B$.  We need to show that
$x=0$.

Case (i): $ B \subset \CC$.  Note that \\
$q(\lambda) : = \langle x, z^\lambda \rangle = \langle \sum c_n u_n, z^n \rangle =
\sum \limits ^N_{n=0} {\bar
c}_n a_n (d_1, \ldots, d_k) \lambda ^n.$\\
So $q$ is a polynomial in $\lambda$ of degree atmost $N$.  Therefore if $q(\lambda)
=0 ~~~ \forall \lambda \in B$, as $ \#B= N+1$, we get $q=0$.  In otherwords $c_n
\equiv 0$, or $x=0$.

 Case (ii): $ \infty \in B$.  Here $\langle x, z^\infty \rangle = C_N =0$.  So
for $\lambda \in \CC, ~~ q(\lambda) = \langle x, z^\lambda \rangle$ is a polynomial
of degree atmost $(N-1)$. Now we can argue as in Case(i). \qed

It is to be noted that ${\cal S}^\bot$ may not contain any unextendible product
basis consisting of orthogonal product vectors.  This can be seen by taking simple
examples such $k =2, d_1 =2, d_2 =3,$ or by making use of beautiful results of Alon
and  Lov\'{a}sz\cite{AL}, where the minimum dimension of an unextendible product basis
consisting of orthogonal vectors is seen to be strictly larger than $(d_1 + \ldots +
d_k -k+1)$ in some special cases.

{\bf Corollary 4}: The minimal dimension of unextendible product bases (of not
necessarily orthogonal vectors)  in ${\cal
H}_1 \otimes \ldots {\cal H}_k$ is $d_1 + d_2 + \ldots +d_k-k+1$.

{\bf Proof}: If $B$ is an unextendible product basis, then $B^\bot$ is completely
entangled.  Hence by Parthasarathy \cite{Pa},  $\dim (B^\bot) \leq d_1 \ldots d_k - (N+1)$, or $\dim (~\mbox {span}~B)\geq N+1.$  Further we have
shown that that there exists an unextendible basis of dimension $(N+1)$. \qed

Now it is a natural question as to what are the possible dimensions of unextendible
bases.  We do not have the answer in general.  But here is the answer when $k=2$.

{\bf Theorem 5:} Suppose $k =2$.  Then for any $m$ with $d_1 +
d_2-1 \leq m \leq d_1d_2$, there exists an unextendible product basis of dimension
$m$.

{\bf Proof :} Consider the decomposition of ${\cal H}$ as ${\cal H} = \bigoplus
\limits ^N_{n =0} {\cal H}^{(n)} = \bigoplus \limits ^N _{n=0} ({\cal S}^{(n)}
\bigoplus {\cal T}^{(n)})$. We know that ${\cal S}$ is completely entangled, so any
subspace of it is also completely entangled.  ${\cal S}^\bot$ has a product basis,
and also by the definition of ${\cal H}^{(n)}, {\cal S}^{(n)} \oplus {\cal
T}^{(n)},$ also has a product basis for every $n$. Therefore ${\cal S}^\bot \cup
{\cal S}^{(n)} = {\cal S}^\bot \cup ({\cal S}^{(n)} \cup {\cal T}^{(n)})$ also has a
product basis. Similarly ${\cal S}^\bot \bigcup (\bigcup \limits _{n \in M} {\cal
S}^{(n)})$ also has a product basis for any subset $M$ of $\{ 0, 1, \ldots, d_1 +
d_2 -1\}$.

As $\dim (\bigcup \limits _{n \in M} {\cal S}^{(n)}) = \sum \limits _{n \in M}
~\mbox{dim}~ ({\cal S}^{(n)})  = \sum
\limits _{n \in M}(a_n (d_1, d_2)-1)$, and the formula for $a_n(d_1, d_2)$ is as in
(1.1), we may choose a suitable set $M$ such that dim $(\bigcup \limits _{n \in M}
{\cal S}^{(n)}) = m -(d_1 +d_2 -1)$.  Now it should be clear that any product basis
of ${\cal S}^\bot \cup (\bigcup \limits _{n \in M} {\cal S}^{(n)})$ does
the job. \qed

Here is an example to show that in general even if a subspace ${\cal M}$ is
completely entangled ${\cal M}^\bot$ may not be spanned by product vectors.  In this
case states with support equal to ${\cal M}^\bot$, as well as states with support
equal to ${\cal M}$ are  entangled!

{\bf Example 2:} Take $k=2, d_1 = d_2 = 4$, and identify ${\cal H}_1 \otimes {\cal
H}_2$ with 4 $\times$ 4 matrices as in Example 1.  Now take
\begin{eqnarray*}
{\cal M} = \{ [a_{ij}]_{0 \leq i , j \leq 3}: 0 &= & a_{00} = a_{01}+a_{10} = a_{02} +
a_{11}+a_{20} = a_{03}+a_{12}\\ & = &a_{21} +a_{30}= a_{13} +a_{22}+a_{31} = a_{23}
+a_{32} =a_{33}  \}
\end{eqnarray*}
Then
\begin{eqnarray*}
{\cal M}^\bot = \{ [a_{ij}]_{0\leq i, j\leq 3}&: &a_{01}= a_{10}, a_{02} = a_{11} =
a_{22},a_{03} =a_{12},\\ & & a_{21} =a_{30}, a_{13} = a_{22} = a_{31}, a_{23}=a_{32}
~~\}
\end{eqnarray*}
It is easily seen that the set of rank one elements in ${\cal M}^\bot$ is given by\\
${\cal R} = \{ c[\lambda^{i+j}]_{0 \leq i, j \leq 3}: c \in \CC, \lambda \in \CC\}
\cup \{c[a_{ij}]: a_{33}=1, a_{ij} =0 \mbox{~otherwise}~\}$ and ${\cal R}$ does not
span ${\cal M}^\bot.$

Note that ${\cal R}^\bot = \{[a_{ij}]: \sum \limits _{i+j=n} a_{ij} = 0 ~~ \forall
n\}= {\cal S}$, which is completely entangled.  So ${\cal R}$ still forms an unextendible
product basis in ${\cal H}_1 \otimes {\cal H}_2$.

{\bf Acknowledgement:} This work was supported by the Department of Science and Technology (India), under
Swarnajayanthi Fellowship Project.


\begin{thebibliography} {99}

\bibitem{HH} M. Horodecki and R. Horodecki, Separability of mixed states : necessary
and sufficient conditions, Phys. Lett. A {\bf 223} (1-2), 1-8, 1996.


\bibitem{VM} D. P. Di Vincenzo, T. Mor, P. W. Shor, J. A. Smolin, B. M. Terhal,
Unextendible product Bases, Uncompletable Product Bases and Bound Entanglement,
Comm. Math. Phys. {\bf 238}, 379-410(2003). quant-ph/9908070.

\bibitem{AL} N. Alon, L. Lov\'{a}sz, Unextendible product bases, preprint,\begin{verbatim}http://www.math.tau.ac.il/~nogaa/PDFS/shor5.pdf.\end{verbatim}


\bibitem{Pa} K. R. Parthasarathy, On the maximal dimension of a completely entangled
subspace for finite level quantum systems, quant-ph/0405077.

\bibitem{Pi} Arthur O. Pittenger, Unextendible product bases and the construction of
inseparable states, Linear Algebra and its Applications {\bf 359}, 235-248(2003).
quant-ph/0208028.


\end{thebibliography}
\end{document}